\documentclass[prl,superscriptaddress,twocolumn,showkeys,amsmath,%
preprintnumbers]{revtex4}
\usepackage{bm} 
\usepackage{graphicx}
\newcommand{\beq}{\begin{equation}}
\newcommand{\eeq}{\end{equation}}
\newcommand{\be}{\begin{eqnarray}}
\newcommand{\ee}{\end{eqnarray}}
\begin{document}
\author{L.~Frankfurt}
\affiliation{School of Physics and Astronomy, Tel Aviv University, 
Tel Aviv, Israel}
\author{M.~Strikman}
\affiliation{Department of Physics, Pennsylvania State University,
University Park, PA 16802, USA}
\author{D.~Treleani}
\affiliation{Department of Physics, University of Trieste INFN and 
ICTP Trieste, I-34014 Trieste, Italy}
\author{C.~Weiss}
\affiliation{Theory Center, Jefferson Lab, Newport News, VA 23606, USA}
\title{Evidence for color fluctuations in the nucleon in
high--energy scattering}
\begin{abstract}
We study quantum fluctuations of the nucleon's parton densities
by combining QCD factorization for hard processes with the notion 
of cross section fluctuations in soft diffraction.
The fluctuations of the small--$x$ gluon density are related to the ratio 
of inelastic and elastic vector meson production in $ep$ scattering. 
A simple dynamical model explains the HERA data and predicts 
the $x$-- and $Q^2$--dependence of the ratio. In $pp/\bar pp$ 
scattering, fluctuations enhance multiple hard processes (but cannot 
explain the Tevatron CDF data), and reduce gap survival in central 
exclusive diffraction.
\end{abstract}
\preprint{JLAB-THY-08-862}
\maketitle
Hard processes in high--energy $ep$ and $pp/\bar p p$ scattering
probe nucleon structure at a resolution scale where it can be described 
in terms of the quark and gluon degrees of freedom of QCD. Essential 
in the analysis of such processes is the method of factorization,
by which the amplitude or cross section is separated into a 
short--distance quark/gluon subprocess, calculable in perturbative 
QCD, and the distributions of the partons in the initial and final hadrons, 
governed by long--distance, non-perturbative interactions. 
Inclusive scattering experiments probe the longitudinal 
momentum densities of the partons. Measurements of hard exclusive 
processes in $ep$ scattering reveal information also about their spatial 
distribution in the transverse plane (generalized parton distributions, 
or GPDs). Such experiments can eventually provide us with a 
full 3--dimensional image of the quark/gluon single--particle 
structure of the nucleon. 

From the perspective of many--body physics the parton densities
represent average characteristics of the nucleon, reflecting the
quantum--mechanical average over configurations in the nucleon 
wave function of different size, number of particles, \textit{etc.}\  
Equally fundamental are the fluctuations around the average value, 
which generally provide information about the nature of the interaction 
in the system. In the case of the quark and gluon densities the 
fluctuations are related to variations of the size and intensity 
of the long--wavelength color fields inside hadrons --- information 
crucial for understanding hadron structure in terms of non-perturbative QCD. 
An interesting question is which experimental observables could
reveal such ``color fluctuations'' inside hadrons.

To see how this problem might be approached, it is worthwhile 
to recall some facts about soft diffractive hadron--hadron scattering 
at high energies and low momentum transfer, $t$. 
In such processes the hadrons can be represented as 
a superposition of states which diagonalize the $T$--matrix and 
experience only individual attenuation (``diffractive eigenstates'').
Elastic diffraction then results from uniform attenuation
of these components, while inelastic diffraction is caused by
different attenuation, which destroys the coherence 
of the original hadronic wave function. This idea was originally formulated
using the language of the eikonal approximation appropriate 
for non-relativistic scattering \cite{Feinberg,Miettinen:1978jb}. 
Later a more general formulation was developed, consistent 
with the fundamental principles of relativistic quantum field theory 
(causality, energy--momentum conservation). It uses the concept of the cross 
section distribution, $P(\sigma )$, which describes the probability 
for the hadrons to scatter in a configuration with given 
cross section \cite{Miettinen:1978jb,Heiselberg:1991is}.
It is normalized as $\int d\sigma \, P(\sigma ) = 1$, and its variance 
is related to the ratio of inelastic and
elastic soft diffractive cross sections at $t = 0$,
\begin{equation}
\omega_{\sigma} \;\; \equiv \;\; 
\frac{\langle\sigma^2\rangle - \langle\sigma\rangle^2}
{\langle\sigma\rangle^2}
\;\; = \;\; 
\left[ \frac{d\sigma_{\text{inel}}}{dt} \! \right/ \! \left.
\frac{d\sigma_{\text{el}}}{dt} \right]_{t=0}^{\text{soft diff}} ,
\end{equation}
where the brackets denote the average with the distribution
$P(\sigma )$. In particular, this formulation allows one to incorporate
color transparency (the vanishing of the interaction 
of small--size configurations) in the $\sigma \rightarrow 0$
behavior of $P(\sigma)$, and the approach to the black--disk regime 
(unitarity limit) at high energies; both are fundamental predictions 
of QCD. Detailed phenomenological studies of various diffractive 
phenomena involving proton and nuclear targets have shown the usefulness 
of this approach and determined the properties of 
$P(\sigma)$ \cite{Blaettel:1993ah,Frankfurt:2000tya}. 
At $pp$ energies $\sqrt{s} \sim 20 \, \text{GeV}$ the variance is 
$\omega_\sigma \sim 0.25$, indicating sizable
fluctuations of the interacting configurations in soft processes.

In this Letter we propose to study color fluctuations in hadrons 
by combining QCD factorization for hard processes with the notion
of cross section fluctuations in soft diffractive processes.
We introduce the concept of a configuration--dependent parton
density and follow its implications for various types of 
high--energy scattering experiments with hard processes. Our investigation
proceeds in three stages. First, we relate the fluctuations of the
gluon density to the ratio of inelastic to elastic hard diffraction
in $ep$ scattering (HERA, future Electron--Ion Collider, or EIC).
Second, we propose a simple model of color fluctuations in the nucleon 
to illustrate and quantify our results. Third, we discuss the
implications of color fluctuations for $pp/\bar p p$ collisions 
with multiple hard processes, and for rapidity gap survival 
in exclusive diffractive $pp$ scattering (Tevatron, LHC). 
The effects described here are not included in present Monte--Carlo 
generators for $pp$ collisions. A more detailed account of our studies 
will be given elsewhere \cite{long}.

Consider diffractive production of vector mesons
in $ep$ scattering at $Q^2 \gtrsim \text{few GeV}^2$, 
$\gamma^\ast_L + p \rightarrow V + X$, where the proton may 
remain intact or dissociate into hadronic states $X$. 
The initial proton state can be expanded in a set of
partonic states characterized by the number of partons and their 
transverse positions, summarily labeled as $|n\rangle$:
$|p \rangle = \sum_n a_n |n\rangle$. Each configuration $n$ has a 
definite gluon density $G(x, Q^2| n)$, given by the expectation
value of the twist--2 gluon operator in the state $|n\rangle$, and the
overall gluon density in the proton is 
\begin{equation}
G(x, Q^2) \;\; = \;\; {\textstyle \sum_n} |a_n|^2 G(x, Q^2| n) \; \equiv \;
\langle G \rangle .
\end{equation}
Because the partonic states appear ``frozen'' on the typical
timescale of the hard scattering process, 
one can use QCD factorization to calculate
the amplitude for vector meson production configuration
by configuration. It is (up to small calculable corrections) 
proportional to the gluon density in that 
configuration \cite{Brodsky:1994kf}. An essential
point is now that in the leading--twist approximation the hard 
scattering process attaches to a single parton, and, moreover,
does not transfer momentum to that parton; it thus does not
change the partonic state $|n\rangle$. Making use of 
the completeness of partonic states, we find that the elastic
($X = p$) and total diffractive ($X$ arbitrary) cross sections are
proportional to
\begin{eqnarray}
(d\sigma_{\text{el}}/dt)_{t=0} &\propto& 
\left[ {\textstyle\sum_n} |a_n|^2 G(x, Q^2| n) \right]^2  \equiv \; 
\langle G \rangle^2 \! , \;\;
\\
(d\sigma_{\text{diff}}/dt)_{t=0} &\propto& 
{\textstyle\sum_n} |a_n|^2 \left[ G(x, Q^2| n) \right]^2  \equiv \; 
\langle G^2 \rangle . \;\;
\end{eqnarray}
For the cross section of inelastic diffraction 
$\sigma_{\text{inel}} = \sigma_{\text{diff}} - \sigma_{\text{el}}$
we thus obtain
\begin{equation}
\omega_g \;\; \equiv \;\; 
\frac{\langle G^2 \rangle - \langle G \rangle^2}{\langle G \rangle^2}
\;\; = \;\; 
\left[ \frac{d\sigma_{\text{inel}}}{dt} \! \right/ \! \left.
\frac{d\sigma_{\text{el}}}{dt} \right]_{t=0}^{\gamma^\ast_L p 
\rightarrow VX} .
\label{omega_g}
\end{equation}
This model--independent relation allows one to infer the fluctuations
of the gluon density from the observable ratio of inelastic and
elastic diffractive vector meson production.
It can be easily generalized to a large variety of hard processes such
as $\gamma^*_L + T \to 2\pi \; \text{(two jets)} + T$, or $\Upsilon$ 
production in ultraperipheral $pp$ collisions at LHC \cite{Baltz:2007kq}. 

Generally, we expect $\omega_g$ to be a weak function of $Q^2$ at 
fixed $x$ (approximate scaling), as the gluon density depends 
only logarithmically on $Q^2$. The $x$--dependence of $\omega_g$ 
at fixed $Q^2$ is difficult to infer from first principles; it depends 
on the ``color flow'' in the nucleon wave function, 
\textit{i.e.}, how the small--$x$ parton densities change with the 
configuration of the large--$x$ constituents. 

%
% FIGURE
%
\begin{figure}
\includegraphics[width=.45\textwidth]{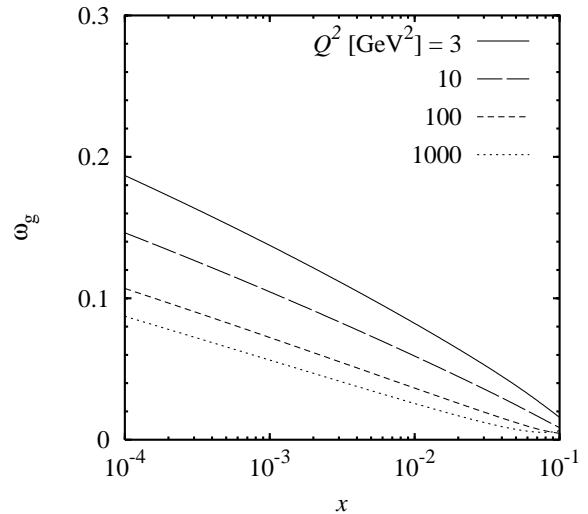}
\caption[]{The variance of fluctuations of the proton's gluon density, 
$\omega_g$, as a function of $x$ for several values of $Q^2$, as obtained
from the scaling model, Eq.~(\ref{g_sigma}), 
and a phenomenological parametrization of the gluon density.}
\label{fig:omg}
\end{figure}
To estimate the variance of the gluon fluctuations at small $x$,
and to study their implications in other hard scattering processes,
we propose here a simple model based on two assumptions: 
(a) At moderate energies 
($\sqrt{s} \sim 20 \, \text{GeV}$) the hadronic cross section of a 
configuration is proportional to the transverse area occupied by 
the color charges in that configuration, 
$\sigma \propto R_{\text{config}}^2$; 
(b) the normalization scale of the parton density changes 
proportionally to the size of the configuration, 
$\mu^2 \propto R_{\text{config}}^{-2} \propto \sigma^{-1}$. 
Assumption (b) is similar to the ``nucleon swelling'' 
model of the EMC effect \cite{Close:1983tn} and implies a simple
scaling relation for the $\sigma$--dependent gluon density:
\begin{eqnarray}
G(x,Q^2 \, | \sigma ) &=& G(x, \xi Q^2), 
\nonumber \\
\xi (Q^2) &\equiv& 
\left( \sigma / \langle \sigma \rangle 
\right)^{\alpha_s (Q_0^2) / \alpha_s (Q^2)} ,
\label{g_sigma}
\end{eqnarray}
where $Q_0^2 \sim 1 \, \textrm{GeV}^2$. Assumption (a) then allows us
to compute the configuration average using the phenomenological 
cross section distribution found in Ref.~\cite{Blaettel:1993ah}.
Figure~\ref{fig:omg} shows the result for the variance of the gluon 
density in this model. At small $x$ and low $Q^2$
it reaches values comparable to the variance of soft 
cross section fluctuations, $\omega$. Note that our evolution--based
model of gluon fluctuations applies primarily to small $x$ 
($\ll 0.1$); at larger $x$ non-perturbative correlations not
included here may become important. Present experimental data on the 
cross section ratio in Eq.~(\ref{omega_g}) are very limited. The value 
$\omega_g \sim 0.15-0.2$ for $Q^2 = 3 \, \text{GeV}^2$ and 
$x \sim 10^{-4} - 10^{-3}$ obtained in our model is consistent 
with the HERA data on vector meson production, where the 
effective scale is $Q^2_{\text{eff}} \sim 2-4 \, \text{GeV}^2$. 
The data also indicate weak dependence of the ratio on $Q^2$ 
and the vector meson mass; however, the limited $Q^2$ range 
and the lack of dedicated studies do not allow us to test 
our model predictions in more detail. Future measurements 
with LHC \cite{Baltz:2007kq} and EIC could significantly 
improve the situation.

\begin{figure}[b]
\includegraphics[width=.45\textwidth]{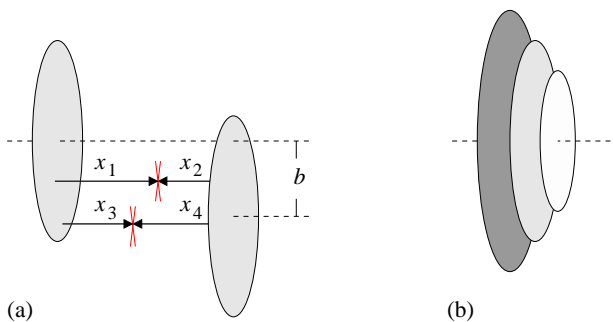}
\caption[]{(a) Double hard scattering in high--energy $pp$--collisions.
(b) Schematic illustration of fluctuations. In the model used here
[\textit{cf}.\ Eq.~(\ref{g_sigma})], configurations
with larger size have larger gluon density.}
\label{fig:dblhard}
\end{figure}

Correlations between fluctuations of the parton densities and
the soft--interaction strength have numerous potential implications 
for high--energy $pp/\bar pp$ collisions with hard processes. 
One example is the relative probability of double
binary parton--parton collisions (see Fig.~\ref{fig:dblhard}a), 
defined as the ratio
\begin{equation}
\frac{\displaystyle
\frac{d\sigma (x_1, x_2; x_3, x_4)}
{d\Omega_{12} \; d\Omega_{34}}}
{\displaystyle
\frac{d\sigma (x_1, x_2)}{d\Omega_{12}} \;
\frac{d\sigma (x_3, x_4)}{d\Omega_{34}}}
= \frac{f(x_1,x_3) f(x_2,x_4)}
{\sigma_{\rm eff} \; f(x_1) f(x_2) f(x_3) f(x_4)} ,
\end{equation}
where $\Omega_{12}$ \textit{etc.}\ are the variables characterizing 
the observed dijets (or photons), and $f(x_1)$ and $f(x_1,x_3)$ 
\textit{etc.}\ are the single and double parton densities, respectively
(we suppress the dependence on the scale). 
The effective cross section $\sigma_{\text{eff}}$ is a measure
of transverse correlation between partons. The FNAL CDF experiment 
measured a value of $\sigma_{\rm eff} = 14.5\pm 1.7^{+ 1.7}_{-2.3} \; 
{\rm mb}$ in $3\text{--jet} + \gamma$ events \cite{Abe:1997bp}, 
a factor of $\sim 3$ smaller than the naive estimate based on a uniform 
distribution of partons with the e.m.\ radius of the nucleon. 
Proper account of the transverse geometry, using gluon GPDs based on
$J/\psi$ photoproduction data, reduces the discrepancy to a factor 
of $\sim 2$ \cite{Frankfurt:2003td}. 
An interesting question is whether fluctuations of the sizes of the 
interacting configurations and the gluon density could explain the 
remaining discrepancy.
%
% FIGURE
%

Elementary geometric arguments show that in the mean--field 
approximation (\textit{i.e.}, no fluctuations)
\begin{equation}
\sigma_{\text{eff}}^{-1} \, \text{(mean field)} \;\; = \;\;
{\textstyle\int} d^2 b \; P_{12} (b) \, P_{34} (b) ,
\label{sigma_eff_mf}
\end{equation}
where $b \equiv |\bm{b}|$ 
is the $pp$ impact parameter, and $P_{12}(b)$ 
describes the $b$--dependence of the probability to have 
two partons with $x_1$ and $x_2$ collide in the $pp$ collision,
\begin{eqnarray}
P_{12} (b) &\equiv& \textstyle {\int d^2\rho_1 \int d^2\rho_2} \; 
\delta^{(2)} (\bm{b} - \bm{\rho}_1 + \bm{\rho}_2 )
\nonumber \\
&\times & F (x_1, \rho_1 ) \; F (x_2, \rho_2 ) ,
\label{P_12}
\end{eqnarray}
with $\int d^2 b \, P_{12}(b) = 1$ (a similar definition applies 
to $P_{34}$). Here $F(x, \rho)$ is the 
normalized transverse spatial distributions of the partons obtained 
from the GPD. To estimate the effect of fluctuations, we now apply the 
same reasoning to the instantaneous 
configurations in the colliding protons, and let both the transverse size 
and the gluon density in these configurations fluctuate as described
by our model [\textit{cf.}\ Eq.~(\ref{g_sigma})]. 
We assume that the transverse size of the configurations 
is proportional to $\sigma$, \textit{i.e.}, the spatial distribution
of partons is $F (x, \rho |\sigma ) = \lambda^{-1} 
F (x, \lambda^{-1/2} \rho)$, 
with $\lambda \equiv \sigma/ \langle \sigma \rangle$ 
(see Fig.~\ref{fig:dblhard}b). With these assumptions we obtain
\begin{eqnarray}
\sigma_{\text{eff}}^{-1} \, \text{(fluct)} &=& 
\left\langle\left\langle \; \phi (x_1 | \sigma_1) \, \phi (x_2 | \sigma_2) \,
\phi (x_3 | \sigma_1) \, \phi (x_4 | \sigma_2) \phantom{{\textstyle \int}}
\right. \right.
\nonumber \\
&\times& \left.\left. 
{\textstyle \int} 
d^2 b \; P_{12} (b|\sigma_1, \sigma_2) \, P_{34} (b|\sigma_1, \sigma_2) 
\; \right\rangle\right\rangle_{12} ,
\nonumber \\
\phi (x_1 | \sigma_1) &\equiv& f (x_1 | \sigma_1) \, / \, 
\langle f (x_1 | \sigma_1) \rangle_1 \;\; \text{etc}., 
\label{sigeff_fluc}
\end{eqnarray}
where the double brackets denote the average over the 
$\sigma$--distributions for both protons. 

Because in multijet production at the Tevatron the typical $x$--values 
of the partons are large ($\sim 0.1$), fluctuations of the parton 
densities are much smaller that in vector meson production at HERA
(see Fig.~\ref{fig:omg}). The dominant effect in Eq.~(\ref{sigeff_fluc}) 
thus comes from fluctuations of the sizes of the interacting configurations.
Keeping only the latter, we obtain a simple analytic result by expanding
in leading order in the variance $\omega_\sigma$:
\begin{equation}
\sigma_{\text{eff}} \, \text{(fluct)}
\;\; = \;\; (1 - \omega_\sigma/2 ) \;\; 
\sigma_{\text{eff}} \, \text{(mean field)};
\end{equation}
numerical studies show that higher--order corrections are negligible 
in practice. One sees that size fluctuations indeed reduce 
$\sigma_{\text{eff}}$, because of the disproportionate enhancement
of multiple hard processes in small--size configurations. 
However, the reduction in our model is found to be only of the order 
$10-15\%$, which cannot account for the discrepancy with the CDF value. 
This indicates that other dynamical mechanisms must be responsible 
for the enhancement of multi--parton collisions, \textit{e.g.}\ 
local transverse correlations between partons as suggested
by a ``constituent quark'' picture of the nucleon \cite{Frankfurt:2004kn}.

We note that much stronger reduction of $\sigma_{\text{eff}}$ was 
obtained \cite{Treleani:2007gi} in a simplified multi--channel 
eikonal model, where all hadronic diffractive states are 
represented by a single channel \cite{Gotsman:2007pn}.
In such a model, to fit the available data on elastic,
inelastic, single and double diffractive cross sections, one needs to
enhance the strength of the Pomeron coupling between diffractive 
eigenstates of smaller radii. In the present model this would
correspond to an increase, rather than a decrease, of the parton
density for small $\sigma$, which is the origin of the different results.

Another class of processes affected by fluctuations is central 
exclusive diffraction, $pp \rightarrow p + H + p$ ($H$ = dijet, 
heavy quarkonium, Higgs boson), widely discussed
as a possible channel for the Higgs boson search at the LHC.
Of interest here is the rapidity gap survival (RGS) probability, $S^2$,
measuring the suppression of the cross section resulting from 
the requirement of no inelastic soft spectator interactions.
This quantity has extensively been studied using models based on 
eikonalized Pomeron exchange \cite{Khoze:2008cx}. 
A recent analysis in a partonic approach with account of the 
transverse geometry (GPDs) found that the suppression of the 
diffractive cross section results mainly from the elimination of 
$pp$ collisions at small impact parameters, in which there is 
a high probability of inelastic interactions \cite{Frankfurt:2006jp}. 
This result was obtained in the mean--field approximation, in which 
the GPDs and the soft--interaction strength are taken at their
average values, with no correlations between them. (The approach of
Ref.~\cite{Khoze:2008cx} gives results comparable to the
mean--field approximation and also ignores such correlations.)

%
% FIGURE
%
\begin{figure}
\includegraphics[width=.33\textwidth]{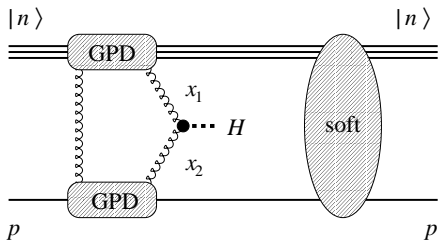}
\caption[]{Fluctuations in central exclusive
diffraction $pp \rightarrow p + H + p$. One of the protons
is treated as a superposition of partonic states $|n\rangle$, 
which undergo diffractive scattering with the other proton
(small fluctuation approximation).}
\label{fig:hardsoft_f}
\end{figure}
To estimate the effects of correlations, we follow the approach 
described above and allow for configuration dependence of both the 
gluon GPDs (\textit{i.e.}, the gluon density and the radius of
the transverse distribution) and the soft--interaction strength. 
The latter we model phenomenologically, using data on the energy dependence 
of $\langle \sigma \rangle$ and $\omega_\sigma$ in soft hadron--hadron 
scattering \cite{long}. At small $x$ the dominant effect comes 
from the correlation of fluctuations of the gluon density with the 
soft--interaction strength. Treating the fluctuations as a small correction 
to the mean--field result we have
\begin{equation}
S^2 ({\text{fluct}}) \;\; = \;\; (1 + 4 \epsilon ) \;
S^2 ({\text{mean field}}) ,
\label{S2_conn}
\end{equation}
where $\epsilon$ is the correction obtained if one of the protons 
fluctuates (see Fig.~\ref{fig:hardsoft_f}); the factor 4 counts 
the number of protons in the diffractive amplitude and its 
complex conjugate. We find $\epsilon \approx -0.07 \, (-0.04)$ for production 
of a system with mass $M_H = 10 \, (100) \, \text{GeV}$ at 
$\sqrt{s} = 2 \, \textrm{TeV}$ (Tevatron), resulting in a reduction of
the RGS probability by a factor $\sim 0.7 \, (0.85)$. The sign
reflects the fact that smaller configurations with higher survival 
probability have a lower density of small--$x$ gluons in 
our model (see Fig.~\ref{fig:dblhard}b). We note that at higher energies 
(LHC) the onset of the black--disk regime in hard interactions causes 
another, more substantial reduction of the RGS probability relative 
to the mean--field result \cite{Frankfurt:2006jp}.

Fluctuations also affect the final--state transverse momentum dependence 
of the cross section for central exclusive diffraction, because they
change the size of the dominant interacting configurations. 
Our findings suggest that the $p_T$ distribution at RHIC and Tevatron 
energies is narrower than given by the mean--field approximation ---
a prediction which could be tested by future measurements of 
diffraction at RHIC.

In sum, the study of quantum--mechanical fluctuations of the quark/gluon 
densities is the natural next step in the exploration of nucleon
structure in QCD, following the mapping of the longitudinal momentum
and transverse spatial distributions of partons. Detailed measurements
of diffractive vector meson production (HERA, EIC) could significantly 
enhance our knowledge of fluctuations of the small--$x$ gluon density. 
They also provide essential input for modeling the dynamics 
of high--energy $pp$ collisions (RHIC, Tevatron, LHC), where fluctuations 
play an important role in multijet production and rapidity gap survival.
How the concept of parton density fluctuations developed here is affected 
by the approach to the unitarity limit at high energies remains an
interesting problem for further study.
\begin{acknowledgments}
Notice: Authored by Jefferson Science Associates, LLC under U.S.\ DOE
Contract No.~DE-AC05-06OR23177. The U.S.\ Government retains a
non-exclusive, paid-up, irrevocable, world-wide license to publish or
reproduce this manuscript for U.S.\ Government purposes.
Supported by other DOE contracts and the Binational Science 
Foundation (BSF).
\end{acknowledgments}
\end{document}